\documentclass[conference,a4]{IEEEtran}
\usepackage{amsmath,amssymb,amsfonts}
\usepackage{mathtools}

\usepackage{stackengine}
\stackMath

\usepackage{algorithmic}
\usepackage{graphicx}
\usepackage{textcomp}
\usepackage{xcolor}
\usepackage{booktabs}
\usepackage{subcaption}
\usepackage{hyperref}

\usepackage{cite}

\usepackage{fancyhdr}

\fancypagestyle{firstpage}{%
\fancyhf{}

 \lhead{This work has been accepted at IEEE/ACM Design, Automation \& Test in Europe Conference (DATE) 2026}
 }

\begin{document}
\fontsize{9.5}{11.5}\selectfont

\bstctlcite{kansystolicarray:BSTcontrol}

\title{
  KAN-SAs: Efficient Acceleration of Kolmogorov-Arnold Networks on Systolic Arrays
}

\author{
   Sohaib Errabii, Olivier Sentieys, Marcello Traiola \\
   University of Rennes, CNRS, Inria, IRISA, Rennes, France
}

\maketitle

\begin{abstract}
Kolmogorov-Arnold Networks (KANs) have garnered significant attention for their promise of improved parameter efficiency and explainability compared to traditional Deep Neural Networks (DNNs).
KANs' key innovation lies in the use of learnable non-linear activation functions, which are parametrized as splines. Splines are expressed as a linear combination of basis functions (B-splines). B-splines prove particularly challenging to accelerate due to their recursive definition.
Systolic Array (SA)-based architectures have shown great promise as DNN accelerators thanks to their energy efficiency and low latency. However, their suitability and efficiency in accelerating KANs have never been assessed.
Thus, in this work, we explore the use of SA architecture to accelerate the KAN inference.
We show that, while SAs can be used to accelerate part of the KAN inference, their utilization can be reduced to 30\%. Hence, we propose KAN-SAs, a novel SA-based accelerator that leverages intrinsic properties of B-splines to enable efficient KAN inference. 
By including a non-recursive B-spline implementation and leveraging the intrinsic KAN sparsity, KAN-SAs enhances conventional SAs, enabling efficient KAN inference, in addition to conventional DNNs.
KAN-SAs achieves up to 100\% SA utilization and up to 50\% clock cycles reduction compared to conventional SAs of equivalent area, as shown by hardware synthesis results on a 28nm FD-SOI technology.
We also evaluate different configurations of the accelerator on various KAN applications, confirming the improved efficiency of KAN inference provided by KAN-SAs\footnote{The source code will be publicly released upon acceptance.}.

\end{abstract}

\thispagestyle{firstpage}

\section{Introduction}

The Kolmogorov-Arnold Network (KAN) is a neural network architecture~\cite{kanpaper} that has garnered interest and been adopted in various applications, such as time series analysis~\cite{vaca-rubio_kolmogorov-arnold_2024}, recommender systems~\cite{park_cfkan}, and medical image segmentation~\cite{li_u-kan_2024}.
The main interest of KANs lies in their improved parameter efficiency and explainability compared to conventional Deep Neural Networks (DNNs).
This is enabled by replacing the conventional scalar weights with learnable spline-based activation functions, which are parameterized in a basis function \textit{(B-spline)}. 
However, this leads to an increase in computational complexity proportional to the size of the basis. Indeed, to compute KAN inference, instead of a single scalar multiply, each function in the basis must first be evaluated at the input, and then a linear combination of all the basis functions is performed. Moreover,
the B-spline functions are evaluated recursively through the Cox-de Boor formula (See Eq. \ref{eq:deboor1}), which makes their acceleration challenging.

Recent artificial intelligence (AI) accelerators increasingly rely on spatial architectures, which have become the \textit{de facto} standard for AI acceleration. Such architectures efficiently execute general matrix multiplication (GEMM), the core operation underlying many AI workloads. Prominent examples include Google’s Tensor Processing Unit (TPU)~\cite{jouppi_datacenter_2017} and NVIDIA’s Tensor Cores, first introduced in the Volta GPU 
microarchitecture~\cite{nvidia_volta}. The efficiency of spatial architectures stems from their ability to maximize
data reuse and minimize data movement, which dominates the energy cost~\cite{horowitz2014computing}.
Building on this foundation, numerous works have proposed further optimizations tailored to specific workloads. For 
instance, Eyeriss~\cite{chen_eyeriss_2017} introduced dataflow optimizations to improve the efficiency of convolutional 
neural networks. Other efforts, such as SCNN~\cite{scnn}, explored computation on compressed weights and activations to exploit zero weights stemming from model pruning and zero activations that occur in ReLU-based networks.
The unique design of KANs limits the effectiveness of existing solutions in accelerating KAN inference compared to other AI applications. Specifically, the recursive nature of B-spline function computation creates a considerable bottleneck, making it challenging to take full advantage of the efficient GEMM executions in spatial architectures such as modern GPUs or Systolic Arrays (SAs)~\cite{kung_why_1982}, which are the foundation of modern TPUs.

Therefore, research efforts have been focusing on new approaches to accelerate KANs (details in Sec.~\ref{sec:related}).
Among recent studies, compute-in-memory (CIM) approaches have been proposed~\cite{kacim,huang_hardware_2025}. In such studies, different ways to approximate the learned non-linear KAN functions or their basis are utilized, such as piece-wise linear (PWL)
approximation.
However, no insights are offered into how spatial (non-CIM)  architectures may be optimized for KANs. 
A recent approach, ArKANe~\cite{arkane}, focuses on the B-spline evaluation bottleneck. It proposes an efficient 
dataflow acceleration methods for the Cox-de Boor recursive formula, achieving a considerable speedup compared to CPU and GPU implementations. 
While all these efforts considerably improve knowledge of KAN acceleration, unfortunately, the current literature lacks solutions to efficiently accelerate KANs on spatial architectures, and particularly on SAs.
To address this gap, this paper analyzes and utilizes the KAN properties to enhance SAs, enabling efficient end-to-end inference acceleration of KAN while maintaining the generality of the accelerator for non-KAN DNN workloads.

As shown in Section~\ref{sec:kansabaseline}, once the B-splines have been evaluated, their linear combination is nothing more than a GEMM operation, which can be accelerated on SAs. 
Regarding the B-spline computation, we observe that a direct floating-point implementation of the recursive evaluation of B-splines is quite costly. For inference-only acceleration, an efficient tabulation strategy of B-splines is possible, thanks to their properties~\cite{huang_hardware_2025}.
Furthermore, B-spline computation results in an $N$:$M$ sparsity pattern in the values processed by the SA, thereby leading to low Processing Element (PE) utilization. Designing a PE of the SA that can handle inputs with a KAN-specific $N$:$M$ sparsity pattern is needed to improve the overall efficiency of the SA~\cite{he_sparse-tpu_2020}.
To the best of our knowledge, there is currently no SA-based accelerator that, in addition to standard DNN workloads, can accelerate KANs by utilizing non-recursive B-spline computation and achieving high PE utilization. In summary, this work makes the following contributions:
\begin{itemize}
    \item We show how a KAN layer can be transformed into a GEMM formulation for execution on a systolic array, and analyze the causes of the inefficiencies that lead to low throughput and poor PE utilization. 
    \item Building upon our analysis, we include in KAN-SAs the needed architectural modifications to handle such inefficiencies. These consist of a PE (i) enhanced with a non-recursive B-spline unit and (ii) providing efficient handling of the $N$:$M$ sparsity pattern generated by B-splines. 
    \item We synthesize KAN-SAs on a 28nm FD-SOI technology and evaluate its improvements compared to a conventional systolic array and to the state-of-the-art approach for B-spline acceleration. 
    Our results show that the proposed KAN-SAs achieves 39.9\% average improvement in PE utilization and 50\% clock cycle reduction on average compared to conventional SAs. Moreover, the B-spline evaluation approach utilized in KAN-SAs achieves more than $72\times$ improvement compared with the state-of-the-art approach.
\end{itemize}

\section{Background}\label{sec:background}

In this section, we present the core operation of a KAN layer and the implications for execution on a systolic array.

\subsection{KAN Layer}
\label{sec:kanlayer}

\begin{figure}[h]
  \begin{center}
\includegraphics[width=0.5\textwidth,trim=0 3 0 3, clip]{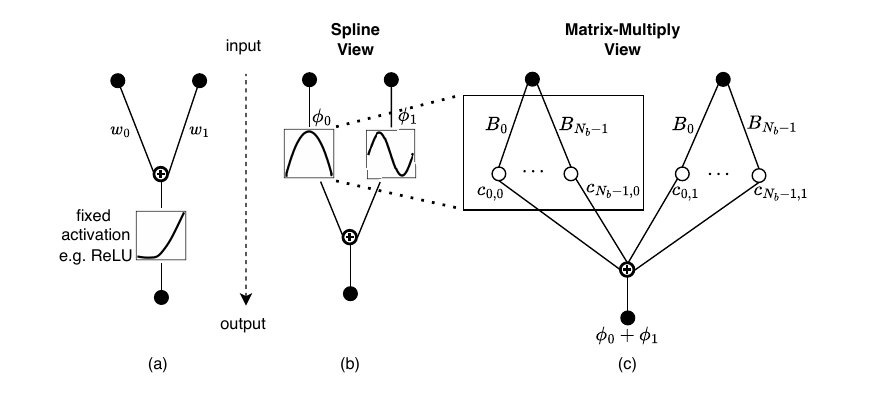}
  \end{center}
  \caption{A single layer of dimensions [2, 1] for MLP (left) and KAN (right).}\label{fig:kanbspview}
\end{figure}

The main difference between a layer in a Multi-Layer Perceptron (MLP) and a KAN layer is illustrated in Fig.~\ref{fig:kanbspview}.
It essentially involves replacing the weights at the connections with learnable activation functions $\phi_i$.

In general, a KAN layer as proposed by~\cite{kanpaper} can be expressed as follows

\begin{equation}
  \label{eq:kanlayer}
  \text{KANLayer}(x) = \sum w_{i}\phi_i(x) + w_b\text{b}(x)
\end{equation}

The second term of the equation, which is not represented in Fig.~\ref{fig:kanbspview} acts as a sort of bias.
It is a typical MLP layer with some fixed non-linear activation applied before the dot product.
It is typically a SiLU but we replace it with a ReLU. 
In the first term, the $w_i$ scales suggested by~\cite{kanpaper} offer better control of the magnitude of the activation functions $\phi_i$; however, at inference time, they can be absorbed in the functions.
While this paper focuses on the first term, the proposed accelerator is capable of executing 
conventional MLP workloads and therefore any KAN layer that follows~\eqref{eq:kanlayer}.

In an MLP layer, a matrix-vector multiplication between the inputs and the weights is followed by an activation using fixed functions such as ReLU.
In the KAN layer, the learned functions are evaluated on the inputs, followed by the sum of all activations (see Fig.~\ref{fig:kanbspview}(b)).

These activation functions are learned through parameterization in some function basis ($\phi(x) 
= \sum_i c_iB_i(x)$). That is, the learnable parameters of the KAN layer are the coefficients $c_i$ for the linear combination in this basis.
Since the parameterization is a linear combination, we can view the KAN layer as a matrix multiply, as shown in  Fig.~\ref{fig:kanbspview}(c). First, all the basis functions are evaluated at the inputs to obtain an intermediate matrix $B$ of dimensions $(M, N_b\times K)$, with $(M, K)$ being the dimensions of the input matrix, and $N_b$ the size of the function basis. The matrix multiplication is then performed with the coefficient matrix of dimensions $(N_b\times K, N)$, with $N$ being the output dimension of the layer (in Fig.~\ref{fig:kanbspview}, $N = 1$, $K = 2$).

Therefore, the second part of the KAN layer can be efficiently accelerated by conventional hardware platforms, such as GPUs and SAs.
However, the first part, i.e., the evaluation of the basis functions, poses a significant computational challenge.
Indeed, the basis first proposed by~\cite{kanpaper} and also adopted by many KAN 
applications~\cite{vaca-rubio_kolmogorov-arnold_2024, kiamari_gkan_2024, bodner_convolutional_2025, li_u-kan_2024}, is the B-spline basis.
This basis is defined by the following Cox-de Boor~\cite{de1972calculating} recursion formula:
\begin{equation} \label{eq:deboor2}
B_{i, P}(x) = \frac{x-t_i}{t_{i + P} - t_i} B_{i, P-1}(x) +\frac{t_{i+P+1}-x}{t_{i+P+1}-t_{i+1}} B_{i+1, P-1}(x)
\end{equation}
with
\begin{equation} \label{eq:deboor1}
B_{i, 0}(x) =
\begin{cases}
1 & \text { if } t_i \leq x<t_{i+1} \\ 0 & \text { otherwise. }
\end{cases}
\end{equation}
Where the points $t_i$ are referred to as knot sequence of the B-spline. $P$ the spline degree and $G$ the grid size of discretization of the input domain 
of the layer. These are hyperparameters of the KAN layer.
Importantly, the recursive formulation of Eq. \ref{eq:deboor1} introduces dependencies between computational stages, making it not efficiently parallelizable on GPUs.

\begin{figure}[htbp]
\includegraphics[width=0.5\textwidth]{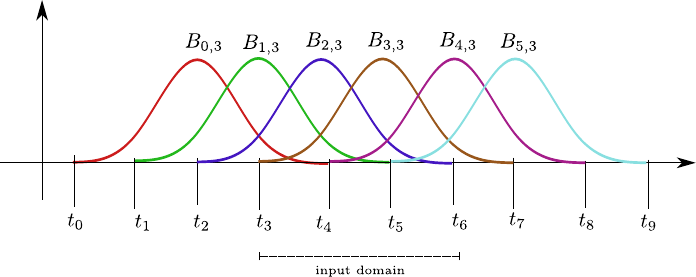}
  \caption{B-spline basis functions of degree $P=3$ on a uniform grid of size $G=3$ for the input domain $[t_3, t_6]$ that is extended on both ends to account for the contributions of B-spline functions that are non-zero within this input domain.}\label{fig:bsplines}
\end{figure}

Fig.~\ref{fig:bsplines} illustrates the case of a uniform grid ($t_{k + 1} - t_{k} = \Delta$) with $G = 3$ and $P = 3$. As shown in the figure, the grid must be extended to account for the contributions of B-splines that are non-zero within the input domain, $P$ extra intervals are added both before and after the input domain, leading to $G + 2P$ total intervals, and $N_b = G + P$ B-spline functions.

\subsection{Related Work}
\label{sec:related}
Regarding the hardware acceleration of KANs, a few approaches have been proposed in the literature.
As for the acceleration of B-spline evaluation, ArKANe~\cite{arkane} has proposed a dataflow acceleration strategy, based on an efficient unroll of the recursive Cox-de Boor formula from Eq.~\ref{eq:deboor2}.
While this can help during training, for inference it incurs significant latency overhead (i.e., several cycles for each B-spline function and input), as it still relies on the recursive computing of B-splines. Moreover, ArKANe approach requires floating-point MAC units and is tightly coupled to AMD-Xilinx FPGA boards, as it relies on Xilinx's recent spatial computing architecture (the Adaptive Intelligent Engine, AIE), for deployment on
AMD-Xilinx Versal ASoCs. Finally, ArKANe does not target the acceleration of the full KAN, but only of B-splines.
For an inference accelerator, a tabulation-based strategy is much more attractive for its low resource cost and minimal
latency. Moreover, it can be adapted for both floating-point and integer-only inference~\cite{integeronly}.

A tabulation-based B-spline implementation was proposed in~\cite{huang_hardware_2025} for a CIM chip based 
on resistive random access memory (RRAM). Unfortunately, besides not targeting systolic arrays, that work does not address the structured sparsity generated by the B-splines. Moreover, that strategy introduces constraints specific to the accelerated software application. For example, grid points were restricted to be powers of two, limiting the acceleration of more generic cases. 
In this work, instead, we aim to implement a more generic systolic-array-based accelerator. The only assumption we make is that of a uniform grid, which does not limit the generality of the B-spline unit. 
Indeed, as demonstrated by~\cite{kanpaper}, it is possible to fine-grain the grid without retraining, using least squares to compute the new coefficients. This enables the approximation of non-uniform grids through finer uniform grids.

Another CIM approach based on CMOS technology was proposed by~\cite{kacim}. This strategy consists of approximating the learned non-linear KAN activations $\phi_i$ of Eq.~\ref{eq:kanlayer} through piece-wise linear (PWL) functions.
In contrast, our approach focuses on accelerating the evaluation of $\phi_i(x)=\sum_i c_iB_i(x)$ on a systolic array architecture through efficient tabular evaluation of B-splines $B_i(x)$ and efficient GEMM operations.

\section{General Principle and Basis Function Unit of KAN-SAs}
\label{sec:kansabaseline}

In this work, we target a weight-stationary systolic array for GEMM acceleration. In our design depicted in Fig.~\ref{fig:kansabaseline}, the weights (i.e., the B-spline coefficients) are pre-loaded into the Processing Elements (PEs) and remain stationary while the input activations are propagated horizontally to other PEs and the resulting partial sums flow vertically towards an accumulator memory. 
This minimizes data movement by reusing weights and activations across processing elements, thereby enabling high energy efficiency.
The memory hierarchy and additional architectural components needed for a practical DNN accelerator are beyond the scope of this paper. 

\subsection{General Architecture of KAN-SAs} \label{sec:kansas}
\begin{figure}[htbp]
\centering
  \includegraphics[width=0.65\columnwidth,trim=26 10 22 10, clip]{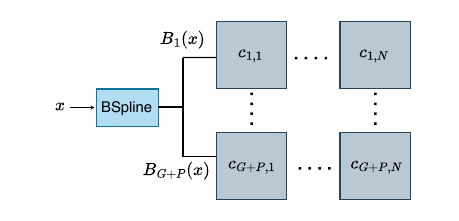}
  \caption{GEMM weight stationary (WS) systolic array with an additional B-spline unit for on-the-fly B-spline activation computation}\label{fig:kansabaseline}
\end{figure}

Following the formulation presented in Section~\ref{sec:kanlayer}, the only requirement for enabling hardware acceleration of KANs through a GEMM systolic array is the generation of the intermediate matrix $B$ of dimensions $(BS, (G + P)K)$, with $BS$ the batch size. $B$ represents the B-spline activations, as shown in Fig.~\ref{fig:kanbspview}(c).
To minimize the data movement required for KAN inference, we aim to enable on-the-fly generation of this intermediate matrix $B$ close to the systolic array.
This can be implemented by adding a basis function unit (the BSpline block in Fig.~\ref{fig:kansabaseline}) that directly streams its values $B_i(x)$ into the systolic array. The rest of this section focuses on this basis function unit, whereas Section \ref{sec:leverage} will detail the PEs of the systolic array.

\subsection{Basis Function Unit} \label{sec:bspline}

Using the Cox-de Boor from Eq.~\ref{eq:deboor2} is not really feasible, as computing a single $P = 3$ function would require $20$ multipliers. Expressing the B-spline in the canonical basis $1, x, \dots, x^3$ is still expensive, considering we need to evaluate the $G + P$ functions. 
Moreover, as mentioned previously, the B-spline functions exhibit several properties enabling an efficient Look Up Table (LUT)-based implementation. 

\subsubsection{Tabulation Strategy}
A B-spline is invariant under a translation and scaling transformation of its knot sequence~\cite{bsplineprops}. i.e.,
\begin{equation*}
    B_{\mathbf{t}, P}(x) = B_{\alpha \mathbf{t} + \beta, P}(\alpha x + \beta), \quad \alpha,\beta \in \mathrm{R}, \quad \alpha \neq 0
\end{equation*}
where $\mathbf{t} = (t_0, \dots, t_{G + 2P + 1})$.
If the grid is uniform, then $t_{k + 1} - t_k = \Delta$. Applying the previous property with $\alpha = \frac{1}{\Delta}$ and $\beta = \frac{-t_0}{\Delta}$, we can map our B-spline defined on an arbitrary knot sequence $t_i$ to one defined on integer knots $0, \dots, G + 2P + 1$, also called cardinal B-spline, such as
\begin{equation*}
B_{\mathbf{t}, P}(x) = B_{[0, \dots, G + 2P + 1], P}(\frac{x - t_0}{\Delta}).
\end{equation*}
Furthermore, with the translation-invariance applied to $B_{k, P}$ we can further write
\begin{equation}\label{eq:alignforcardinal}
B_{t_k, P}(x) = B_{k, P}(\frac{x - t_0}{\Delta}) = B_{0, P}(\frac{x - t_0}{\Delta} - k).
\end{equation}
This makes the B-spline independent of the KAN layer grid points $t_i$, since we only need to tabulate the function $B_{0, P}$, thereby enabling a ROM-based implementation of the B-spline unit.
Of course, the unit must still perform the alignment required for $x$ shown in Eq.~\ref{eq:alignforcardinal}.

More importantly, the key property of B-splines for our purposes is their local support, which is given by the extreme knots used in their definition of Eq.~\ref{eq:deboor1}, $B_{t_k, P}(x) = 0, \quad x \notin [t_k, t_{k + P + 1}[$. This is also shown clearly in Fig.~\ref{fig:bsplines}, and it implies that for any input $x$, there is at most $P + 1$ non-zero B-spline activations.

To summarize, the B-spline unit must first handle the alignment required by Eq.~\ref{eq:alignforcardinal}, which enables the use of stored $B_{0, P}$ values. And then, relying on translation-invariance, it must infer the value of all $P + 1$ non-zero B-splines.
Moreover, to efficiently tabulate $B_{0, P}$, we must leverage the fact that the cardinal B-spline is symmetric with respect to the midpoint of the support (for $B_{0, P}$, that is $[0, \dots, P + 1]$), $\frac{P + 1}{2}$~\cite{bsplineprops}.
Therefore, we only need to store half the B-spline corresponding to the interval $[0, \frac{P + 1}{2}]$. For instance, for $P = 3$, we only need to sample points from the interval $[0, 2]$.

\begin{figure}[!t]
    \centering
\includegraphics[width=.85\columnwidth,trim=0 0 5 5, clip]{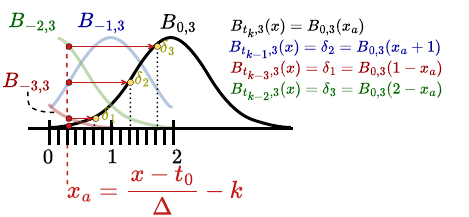}
    \caption{
    Illustration of efficient tabular storing of half a cubic B-spline and computation of the others by aligning and shifting the input $x$. 
    Only the values of $B_{0, 3}$ corresponding to the black sample points on the horizontal axis quantizing the interval $[0, 2]$ are stored. 
    For a given input $x$, all the B-spline values can be obtained by retrieving $B_{0, 3}$ stored values at the aligned input $x_a$ and by shifting the evaluation point. E.g., $B_{t_{k - 2},3}(x) = B_{0,3}(2 - x_a)$.
    }
    \label{fig:bsplinelutstrat}
\end{figure}

The example in Fig.~\ref{fig:bsplinelutstrat} helps clarify the approach. For an input $x$, only $P + 1 = 4$ non-zero B-splines exist: 
$B_{t_{k - 3},3}(x), B_{t_{k - 2},3}(x), B_{t_{k - 1}, 3}(x),B_{t_k,3}(x)$. The aligned $x_a$ input shown in the figure ensures that these values correspond respectively to $B_{-3, P}(x_{a}), B_{-2, 3}(x_{a}), B_{-1, 3}(x_{a}), B_{0, 3}(x_{a})$.

\subsubsection{Implementation of the BSpline Unit}

\begin{figure}
  \centering
  \includegraphics[width=0.48\textwidth]{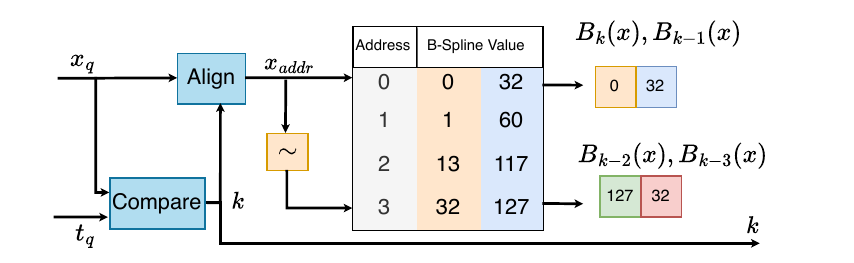}
  \caption{
  Simplified version of the B-Spline unit. The address is inverted, and the corresponding
  values are reverse-packed to obtain the remaining non-zero activations. 
  The example values $0, 32$ at the output correspond to $x_{addr} = 0$. The unit also returns the values at $\sim x_{addr} = 3$ in reverse. i.e. $127, 32$.
  }
  \label{fig:bsplinelut}
\end{figure}

Fig.~\ref{fig:bsplinelut} shows the table storing the B-spline values of $B_{0, P}$ required to infer all the non-zero B-spline 
activations. The \textit{Align} unit implements the previously mentioned alignment in Eq.~\ref{eq:alignforcardinal} and the 
\textit{Compare} unit performs an interval search to compute $k$, necessary for both the alignment as well as the systolic array, as will be shown in Sec.~\ref{sec:leverage}. 
The LUT address for retrieving the stored value for $x$ is the quantized aligned input $x_a \in [0, 1]$ to the integer range $[0, 255]$,

The unit must obtain $x_{addr}$ using integer arithmetic based on its quantized 
inputs $x_q$ and $\mathbf{t_q}$. Using affine integer quantization scheme with 
Eq.~\ref{eq:alignforcardinal} we obtain the LUT address computed in \textit{Align} unit as
\begin{equation}\label{eq:lutaddr}
x_{addr} = \text{clip}\left( (G + 2P)(x_q - t_{q0}) - 255\times k, 0, 255 \right).
\end{equation}
Since we use an address based on $x_a \in [0, 1]$ and we need to store values in $[0, 2]$. We simply 
store the value corresponding to $x_a + 1$ in the same address as $x_a$. This value corresponds to $B_{t_{k - 1}, 3}(x)$ as shown in Fig.~\ref{fig:bsplinelutstrat}, hence the two values per row in Fig.~\ref{fig:bsplinelut}. For the other values $B_{t_{k - 3},3}(x), B_{t_{k - 2},3}(x)$, which correspond to $B_{0, 3}(1 - x_a)$ and $B_{0, 3}(1 - x_a + 1)$, they are stored in the address corresponding to $1 - x_a$, i.e., $255 - x_{addr}$. This is implemented in the inversion unit \textit{$\sim$} in Fig.~\ref{fig:bsplinelut}.

\section{Leveraging B-splines Properties in KAN-SAs Processing Elements} \label{sec:leverage}

In this section, we present the analysis of the B-spline properties presented previously, and how they enable the efficient acceleration 
of KAN networks on SA-based accelerators.
In particular, we show how to provide efficient handling of the $N$:$M$ sparsity pattern generated by B-splines.

\subsection{On the Inherent Sparsity of B-splines}
As illustrated in Fig.~\ref{fig:bsplines} and explained in Section~\ref{sec:bspline}, B-splines have local support, i.e., the B-spline $B_{t_k, P}$ is only non-zero
within $[t_k, t_{k + P + 1}]$. Therefore, if $x \in [t_k, t_{k + 1}]$, then 
only the $P + 1$ functions $B_{t_{k - P}}(x),\dots, B_{t_k}(x)$ are non-zero among all $G + P$ functions. 
Moreover, when $x$ is within the grid extension ($k < P$ or $k > G + P - 1$), there are even fewer non-zero B-splines. For example, with $G = 10, P = 3$, each input would at most contribute $4$ non-zero B-spline activations among $13$, which leads to at most $\frac{4}{13}\approx30\%$ PE utilization (i.e., computations involving non-zero B-spline activations) in the SA (as shown later in Fig.~\ref{fig:util_app}).

To address this inefficiency, the B-spline unit must output only the $ P+1$ contiguous non-zero activations along with the integer index $k$ specifying their positions among all the $ G+P$ B-splines. 
This nice property of B-splines enables the use of an $N$:$M$ structured sparsity-aware PE (where $N=P + 1$, $M=G + P$) to avoid useless multiplications with zero.
This means that while the classical scalar PE in Fig.~\ref{fig:kansabaseline} performs $psum_i + c_iB_i(x)$, a KAN-optimized PE has to be vectorized to perform $psum_i + \sum_{i = 0}^{P}c_{k - i}B_{k - i}(x)$, i.e. only involving $B_{k - i}(x),\ i\in \left[ 0,P\right]$ non-zero B-spline values.
This $N$:$M$ structured pattern, also known as density bound block (DBB), has been previously enforced in conventional DNNs static weights~\cite{dbb,Kang_2020,sparsetc} through ad-hoc pruning. 
Conversely, in the KAN case, this $N$:$M$ sparsity appears dynamically in the B-spline activations, and it is guaranteed by their local support property.

\subsection{Designing the Processing Elements of KAN-SAs}\label{sec:pe}

A conventional systolic array uses a scalar processing element performing a multiply-accumulate.
To harness the benefit of the intrinsic  B-spline sparsity, we design an $N$:$M$ sparsity-aware vector processing element performing a multiply-accumulate between the $N$ non-zero B-spline values and their corresponding 
coefficients among all the $M$ coefficients.

\begin{figure}
  \centering
  \includegraphics[width=0.6\columnwidth, trim=5 4 5 22, clip]{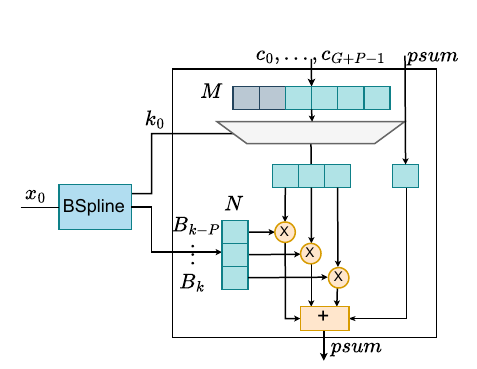}
  \caption{Proposed KAN-SAs PE architecture showing how the B-spline unit interacts with the $N$:$M$ vector PE. Different from the scalar PE SA in Fig.~\ref{fig:kansabaseline}, each B-spline unit streams its output to a single row of $N$:$M$ PEs.}
  \label{fig:kanoptsa}
  \vspace{-0.5cm}
\end{figure}

This is shown in the Fig.~\ref{fig:kanoptsa}, where each B-spline unit sends the $P + 1$ non-zero activations to its corresponding
row in the array along with in the index $k_0$ as a control signal for the multiplexer selecting the right $N = P + 1$ coefficients among 
$M = G + P$. Therefore, while the $N$ multiplications occur in parallel, the additional $M$-to-$N$ multiplexer and multi-operands adder for summing the partial products do increase the delay of the critical path.
Similar to the activations, the indices $k$ are propagated and reused horizontally along the array.

\section{Evaluation} \label{sec:eval}

In this section, we evaluate the proposed KAN-SAs architecture against the conventional SA setup.
In the conventional SA, we assume B-spline units feeding a systolic array with scalar PEs, not able to handle the intrinsic $N$:$M$ B-spline sparsity.
We synthesized the conventional SA solution and the proposed one with Synopsys Design Compiler v2025.6, targeting the ST 28nm FD-SOI PDK~\cite{cathelin_fully_2017}.
The integer-only implementation, quantized as proposed by~\cite{integeronly}, was validated with a software baseline. We observe less than $1\%$ in accuracy drops for all the models (e.g., MNIST-KAN drops from $96.58\%$ to $96.0\%$).

\subsection{PE Comparison}\label{sec:results}

In this section, we compare the hardware synthesis results for the scalar conventional PE and the $N$:$M$ PE, as provided in Table \ref{tab:pe_pdp}. Post-synthesis power estimation was carried out using activity-based analysis at a frequency of $500$MHz.
We first study the delay for different sparsity parameters $N$ and $M$. As mentioned in Section~\ref{sec:pe}, there is an increase in the delay of the critical path due to both the $M$-to-$N$ multiplexer and the 
$N + 1$-operands adder. For instance, when increasing $N$
from 2:6 to 4:6 in Table~\ref{tab:pe_pdp}, the increase in delay is due to the adder since the multiplexing occurs in 
parallel. Similarly, when increasing $M$, the adder is not affected, but the multiplexer's delay increases.
As a consequence of the added logic, as shown in the table, the power of $N$:$M$ PEs also increases with respect to 1:1. The switching activity was extracted from simulation traces for a typical KAN workload.
The coefficients are loaded in the PE and then reused for several cycles with different 
activations.
However, the normalized energy reported in the table shows the advantage of the $N$:$M$ PE, as a KAN workload running on an $N$:$M$ PE  array takes $(G + P)$ times fewer cycles than a scalar PE, with $N=P+1,\ M= G + P$.
The energy is estimated based on the reported power in the table and the number of cycles a KAN workload requires to run on a scalar PE by multiplying the power in Table~\ref{tab:pe_pdp} by the number of cycles needed to run a typical KAN workload. The 1:1 PE takes $(G+P)$ times more cycles than a PE handling $N$:$M$ sparsity. E.g., for $P=G=3$ (sparsity 4:6), the 1:1 PE will need $6\times$ more cycles than 4:6.

\begin{table}[t]
  \centering
  \caption{
  ST28nm FD-SOI post-synthesis delay and power  
  for 8-bit inputs and 32-bit output PE at a target frequency of $500$ MHz and estimated normalized energy.
  Columns refer to the sparsity pattern $N$:$M$; 1:1 represents the scalar PE.
  }
  \label{tab:pe_pdp}
    \centering
  \begin{tabular}{c|c|c|c|c|c|c}
    N:M & {1:1} & {1:2} & {2:4} & {2:6} & {4:6} & {4:8} \\ \hline
Delay (ns) & {1.02} & {1.05} & {1.15} & {1.19} & {1.28} & {1.31} \\ \hline
Power (mW) & 0.35 & 0.40 & 0.62 & 0.77 & 0.98 & 1.12 \\ \hline
\begin{tabular}[c]{@{}l@{}}Normalized\\ Energy\end{tabular} & 1.00 & 0.57 & 0.44 & 0.37 & 0.47 & 0.40 \\\hline
    \end{tabular}
    \vspace{0.5em}
  
  \vspace{-7pt}
\end{table}

\subsection{B-spline Acceleration Comparison with Previous Work}

As already mentioned, ArKANe~\cite{arkane} proposes an acceleration of the recursive implementation of B-splines. A direct comparison is not straightforward, as ArKANe B-spline implementation evaluates the B-splines
using floating-point arithmetic across multiple cycles for the objective of KAN training acceleration. Moreover, the reported results correspond to a mapping on Xilinx FPGA over multiple AIE tiles.
However, to propose a fair comparison, we estimate the possible area consumption for ArKANe proposal and compare it to KAN-SAs B-spline unit. 
ArKANe wavefront algorithm requires $P + 1$ processing elements and  
has a latency of $(P + 1)\times \text{PE}_{latency}$ cycles to evaluate a specific B-spline $B_{i, P}$. Then, thanks to pipelining, it requires 
$G + P - 1$ cycles for all $G + P$ activations. Hence, we can compute the ArKANe number of cycles for $M$ inputs as $(P+1)PE_{latency}+G+P-1+M$. 
We consider an existing single-precision FMA implementation, FPMax~\cite{fpmax}, as a reference for the ArKANe FP$32$ multiply-accumulate (FMA) operation in each PE. In FPMax, a single-precision FMA circuit has $\text{PE}_{latency} = 4$ and is estimated to occupy approximately $0.0081 mm^2$ of standard cell area.
By contrast, our tabulation-based B-spline unit occupies $450 \mu m^2$, and requires at most a single cycle to retrieve the values of all $G + P$ B-splines for a certain input.
Therefore, in the same estimated area for ArKANe, i.e., $4\times 0.0081 mm^2$ we can fit $72$ B-spline units to feed $72$ rows of the systolic array.
Tabulating the B-splines can offer a minimum of $72\times$ speedup for high values of M over the recursive floating-point implementation.

\subsection{KAN Applications Benchmark}

\begin{figure}[htbp]
\centering
  \begin{subfigure}[b]{0.46\columnwidth}
  \includegraphics[width=\columnwidth,trim=3 3 3 3, clip]{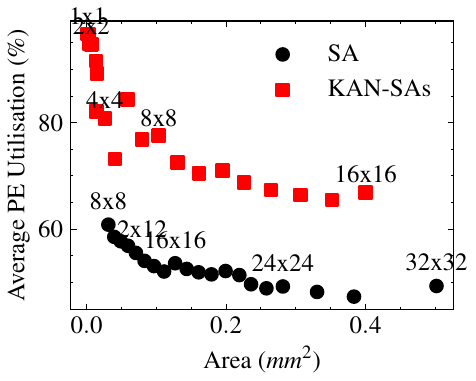}
  \caption{\label{fig:util_area_st28}}
  \end{subfigure}
\begin{subfigure}[b]{0.51\columnwidth}
  \centering
\includegraphics[width=\columnwidth,trim=5 3 4 2, clip]{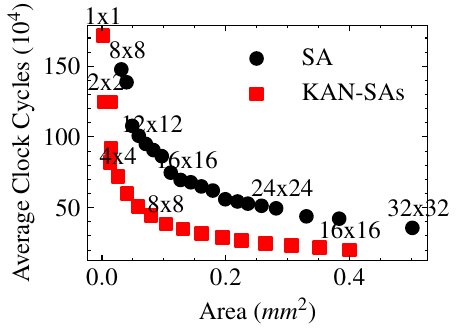}
  \caption{\label{fig:runtime_area_st28}}
\end{subfigure}
\caption{Average PE utilization (a) and runtime in clock cycles (b) across all applications vs. the area obtained post-synthesis for both conventional SA and KAN-SAs, for various sizes of the SA ($R\times C$).}
\end{figure}

We perform a design space exploration of different configurations of the proposed architecture on a representative set of KAN applications collected from prior work and reported in Table~\ref{tab:applications}.
Each of the applications contributes a certain number of KAN workloads, i.e., matrix multiplications where the left matrix is B-spline activations $B$ mentioned in Section~\ref{sec:kanlayer} and the right matrix is the coefficients. For instance, the application Catch22-KAN relies on a single KAN layer $[22, X]$ where $X$ is the number of classes for one of the UCR datasets~\cite{ucrtsdata}. This layer implies a matrix $B$ of dimensions $(BS, 22\times(G + P))$. Some of these workloads also include the MLP bias term mentioned in Section~\ref{sec:kanlayer}, which is included in the evaluation.
\begin{table}[hbtp]
  \caption{Collected KAN workloads from prior work.
  $X$ in CF-KAN takes values from $[2810, 34395, 6969]$. In Catch22-KAN, $X$ is the number of classes in UCR time series dataset, and we consider it smaller than $60$ in our experiments.
  }
  \label{tab:applications}
  \begin{tabular}{cccc}
    \toprule
    Application & Layers & G & P \\
    \midrule 
   5G-STARDUST~\cite{vaca-rubio_kolmogorov-arnold_2024} & [168, 40, 40, 40, 24] & 5 & 3 \\
   Catch22-KAN~\cite{Ismail-Fawaz2023kan-c22-4-tsc} & [22, X] & 3 & 3 \\
   CF-KAN~\cite{park_cfkan} & [X, 512, X] & 2 & 3 \\
   U-KAN~\cite{li_u-kan_2024} & [512, 1024, 512], [512, 512] & 5 & 3 \\
   GKAN~\cite{kiamari_gkan_2024} & [200, 16, 7], [100, 20, 7] & 2,3 & 1,2,3 \\
   Prefetcher~\cite{prefetching} & [5, 64, 128] & 4 & 3 \\
   MNIST-KAN~\cite{efficientkan} &  [784, 64, 10] & 10 & 3 \\
   ResKAN18~\cite{convkan} & 20 ConvKAN layers & 3 & 3 \\
  \bottomrule
\end{tabular}
\vspace{-7pt}
\end{table}
Some studies explore different values of $G$ and $P$. However, we limit our evaluation to workloads with $P \le 3$.
The applications MNIST-KAN and ResKAN18 are models we implemented 
and trained on MNIST~\cite{mnist} and  CIFAR10~\cite{cifar10}, respectively. MNIST-KAN is a two-layer KAN $[784, 64, 10]$,
and the ResKAN18 is a ResNet18 architecture, where the scalar filter weights in convolutions are replaced by learnable splines (referred to as ConvKAN or KANConv in prior works~\cite{bodner_convolutional_2025,drokin2024convkan}). 
The average results reported in Figs.~\ref{fig:util_area_st28} and~\ref{fig:runtime_area_st28} consider different dimensions $R$ rows and $C$ columns for the array. In the graphs, for reference, we mark some points that correspond to square SAs (e.g., 2x2, 4x4, etc.). The other parameters are fixed as int$8$ multiplication and int$32$ accumulation, $G = 5$ and $P = 3$.
Hence, results are averaged over all collected workloads except MNIST-KAN, as it requires $G = 10$.
Since the main objective of the study is to evaluate the proposed improvements, we focus solely on B-spline sparsity without considering other dynamic sources of sparsity, such as zero coefficients (i.e., weights) or activations.
The workloads are tiled for running on the weight-stationary SA. 

\begin{figure}[htbp]
\centering
\includegraphics[width=0.75\columnwidth,trim=3 15 3 3, clip]{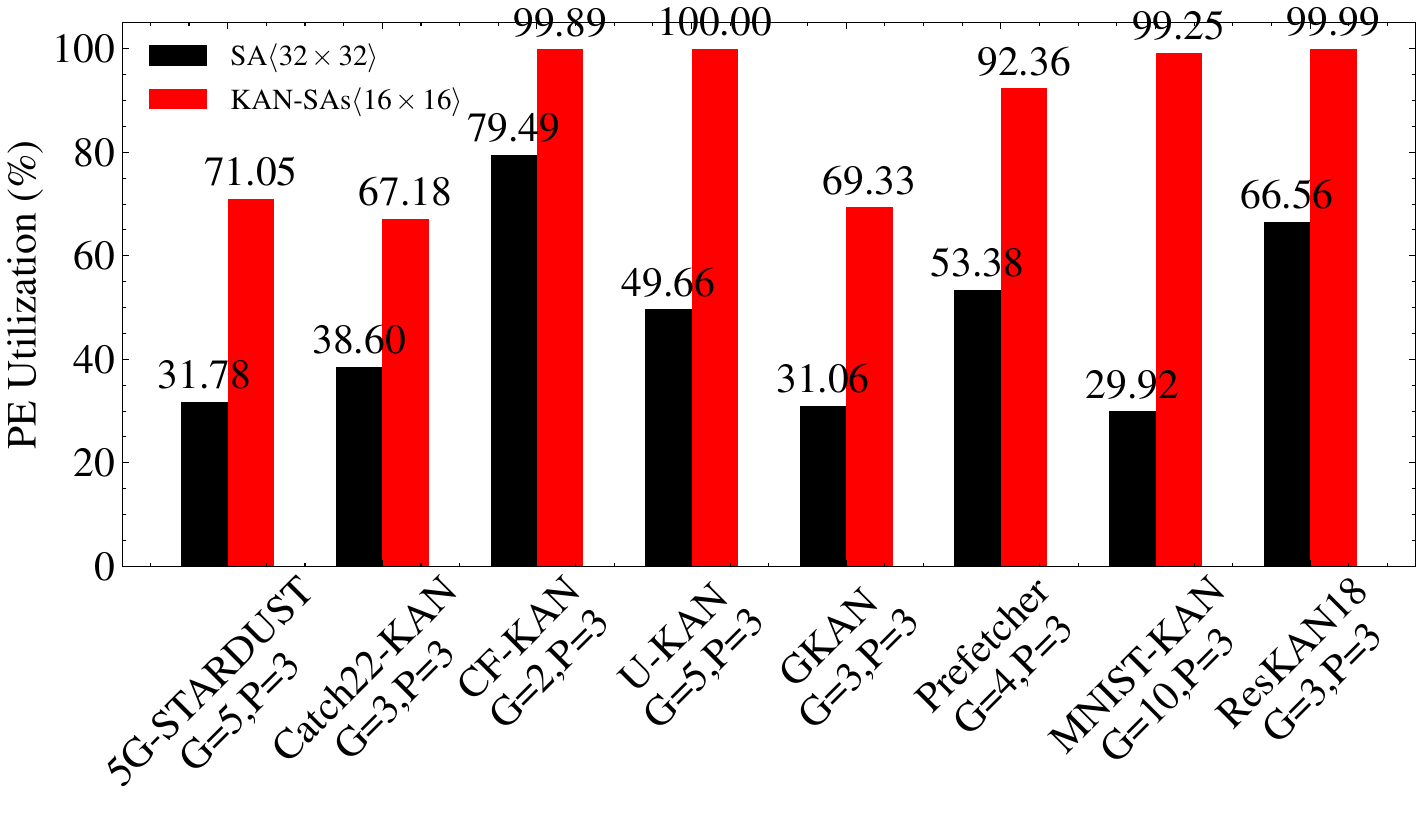}
  \caption{PE Utilization(\%) for KAN workloads.}\label{fig:util_app}
  \vspace{-7pt}
\end{figure}

In Fig.~\ref{fig:util_area_st28}, we plot, for both conventional SA and KAN-SAs, the average PE utilization across all applications vs. the area obtained post-synthesis, for various sizes of the SA ($R\times C$).
The figure clearly shows that the proposed solution consistently improves the average PE utilisation of the SA for different design parameters. However, it can also be noticed that, while KAN-SAs consistently provides PE utilization $>65\%$, there is still some effect preventing 100\% utilization.
The PE utilization is fundamentally impacted by two distinct effects. 
The first and most important one is addressed by KAN-SAs, which is the \textit{sparsity generated by the B-splines}. 
The second is \textit{imperfect tiling}, i.e., some workloads' dimensions are not a multiple of the SA dimensions, and is intrinsic to the workload being executed.
To give more details, in Fig.~\ref{fig:util_app}, we show the per-application average PE utilization for specific configurations of the arrays with similar areas, i.e., $0.47 mm^2$ for KAN-SAs $16\times16$ and $0.50mm^2$ for $32\times32$ scalar PE SA. 
Since KANs are still in their early stages, most applications have small dimensions. This leads to a noticeable imperfect tiling effect in these applications. 
Also, the values of G impact the utilization. When $G$ has high values, two effects are present: (i) the overall number of parameters also increases, and (ii) the sparsity grows, as shown in Sec.~\ref {sec:leverage}. The first effect intrinsically reduces the imperfect tiling, and the second reduces the utilization in conventional SA. 
For instance, MNIST-KAN with $G = 10$ has low utilization ($30\%$) when executed on SA because of the high G value. Conversely, on KAN-SAs, it has a high utilization rate (99.25\%) because the high G value leads to a large number of parameters, thereby reducing the impact of imperfect tiling.
As can be seen from Table~\ref{tab:applications}, ResKAN18, CF-KAN, U-KAN, and Prefetcher applications are large compared to the others, which reduces the imperfect tiling impact, even for low G values. Additionally, low G values result in higher PE utilization, also for conventional SA. The absolute utilization improvement of KAN-SAs over conventional SA is 39.9\% on average over all the applications, with a maximum of 69.3\% for MNIST-KAN.

Finally, Fig.~\ref{fig:runtime_area_st28} shows, for both conventional SA and KAN-SAs, the average runtime (in clock cycles) across all applications vs. the area.
KAN-SAs consistently provides reduced runtime by 2$\times$ compared to conventional SA at the same area and reduced area by 2$\times$ at the same runtime.
The improvement stems from the fact that while the scalar PEs in the conventional SA can load weight tiles of size $R\times C$, the $N$:$M$ PEs in KAN-SAs load $(R\times M, C)$ tiles of B-spline activations and $(R\times N, C)$ tiles of non-KAN workloads (i.e., the MLP term in Eq.~\ref{eq:kanlayer}), leading to faster execution.

\section{Conclusion}

In this paper, we presented KAN-SAs, the first systolic-array-based efficient hardware accelerator for KANs. KAN-SAs efficiency is enabled by a thorough analysis of the inefficiencies that the B-spline functions inherently introduce, i.e., recursive nature and sparsity. KAN-SAs addresses them through an enhanced Processing Element including a non-recursive B-spline unit and efficient handling of B-spline sparsity.
Results show that, compared to conventional SAs, KAN-SAs achieves 39.9\% average PE utilization improvement, 50\% average clock cycle reduction, and 72$\times$ speedup in B-spline evaluation compared to the state-of-the-art approach.

\section*{Acknowledgment}
This work was supported by the French National Research Agency (ANR) through the RADYAL project ANR-23-IAS3-0002.

\bibliography{kansystolicarray.bib}

\begin{thebibliography}{10}
\providecommand{\url}[1]{#1}
\csname url@samestyle\endcsname
\providecommand{\newblock}{\relax}
\providecommand{\bibinfo}[2]{#2}
\providecommand{\BIBentrySTDinterwordspacing}{\spaceskip=0pt\relax}
\providecommand{\BIBentryALTinterwordstretchfactor}{4}
\providecommand{\BIBentryALTinterwordspacing}{\spaceskip=\fontdimen2\font plus
\BIBentryALTinterwordstretchfactor\fontdimen3\font minus
  \fontdimen4\font\relax}
\providecommand{\BIBforeignlanguage}[2]{{%
\expandafter\ifx\csname l@#1\endcsname\relax
\typeout{** WARNING: IEEEtran.bst: No hyphenation pattern has been}%
\typeout{** loaded for the language `#1'. Using the pattern for}%
\typeout{** the default language instead.}%
\else
\language=\csname l@#1\endcsname
\fi
#2}}
\providecommand{\BIBdecl}{\relax}
\BIBdecl

\bibitem{kanpaper}
Z.~Liu, Y.~Wang, S.~Vaidya, F.~Ruehle, J.~Halverson \emph{et~al.}, ``Kan:
  Kolmogorov-arnold networks,'' 2024, arXiv:2404.19756 [cs.LG].

\bibitem{vaca-rubio_kolmogorov-arnold_2024}
C.~J. Vaca-Rubio, L.~Blanco, R.~Pereira, and M.~Caus, ``Kolmogorov-{Arnold}
  {Networks} ({KANs}) for {Time} {Series} {Analysis},'' Sep. 2024,
  arXiv:2405.08790 [eess.SP].

\bibitem{park_cfkan}
J.-D. Park, K.-M. Kim, and W.-Y. Shin, ``Cf-kan: Kolmogorov-arnold
  network-based collaborative filtering to mitigate catastrophic forgetting in
  recommender systems,'' 2024, arXiv:2409.05878 [cs.IR].

\bibitem{li_u-kan_2024}
C.~Li, X.~Liu, W.~Li, C.~Wang, H.~Liu \emph{et~al.}, ``U-{KAN} {Makes} {Strong}
  {Backbone} for {Medical} {Image} {Segmentation} and {Generation},'' Aug.
  2024, arXiv:2406.02918 [eess.IV].

\bibitem{jouppi_datacenter_2017}
N.~P. Jouppi, C.~Young, N.~Patil, D.~Patterson, G.~Agrawal \emph{et~al.},
  ``In-{Datacenter} {Performance} {Analysis} of a {Tensor} {Processing}
  {Unit},'' in \emph{ACM/IEEE 44th Annual International Symposium on Computer
  Architecture (ISCA)}, 2017, pp. 1--12.

\bibitem{nvidia_volta}
``Nvidia tesla v100 gpu architecture: The world's most advanced data center
  gpu,'' Whitepaper,
  \url{https://images.nvidia.com/content/volta-architecture/pdf/volta-architecture-whitepaper.pdf},
  2017.

\bibitem{horowitz2014computing}
M.~Horowitz, ``1.1 computing's energy problem (and what we can do about it),''
  in \emph{2014 IEEE International Solid-State Circuits Conference Digest of
  Technical Papers (ISSCC)}, 2014, pp. 10--14.

\bibitem{chen_eyeriss_2017}
Y.-H. Chen, T.~Krishna, J.~S. Emer, and V.~Sze,
  ``\BIBforeignlanguage{en}{Eyeriss: {An} {Energy}-{Efficient} {Reconfigurable}
  {Accelerator} for {Deep} {Convolutional} {Neural} {Networks}},''
  \emph{\BIBforeignlanguage{en}{IEEE Journal of Solid-State Circuits}},
  vol.~52, no.~1, pp. 127--138, Jan. 2017.

\bibitem{scnn}
A.~Parashar, M.~Rhu, A.~Mukkara, A.~Puglielli, R.~Venkatesan \emph{et~al.},
  ``Scnn: An accelerator for compressed-sparse convolutional neural networks,''
  2017, arXiv:1708.0448 [cs.NE].

\bibitem{kung_why_1982}
H.~T. Kung, ``Why systolic architectures?'' \emph{Computer}, vol.~15, no.~1,
  pp. 37--46, Jan. 1982.

\bibitem{kacim}
C.~Sudarshan, P.~Manea, and J.~P. Strachan, ``A {Kolmogorov}–{Arnold}
  {Compute}-in-{Memory} ({KA}-{CIM}) {Hardware} {Accelerator} with {High}
  {Energy} {Efficiency} and {Flexibility},'' Jan. 2025, preprint
  10.21203/rs.3.rs-5804189/v1.

\bibitem{huang_hardware_2025}
W.-H. Huang, J.~Jia, Y.~Kong, F.~Waqar, T.-H. Wen \emph{et~al.},
  ``\BIBforeignlanguage{en}{Hardware {Acceleration} of {Kolmogorov}-{Arnold}
  {Network} ({KAN}) for {Lightweight} {Edge} {Inference}},'' in
  \emph{\BIBforeignlanguage{en}{30th {Asia} and {South} {Pacific} {Design}
  {Automation} {Conference}}}.\hskip 1em plus 0.5em minus 0.4em\relax ACM, Jan.
  2025, pp. 693--699.

\bibitem{arkane}
Y.~Wu and M.~T. Arafin, ``Arkane: Accelerating kolmogorov-arnold networks on
  reconfigurable spatial architectures,'' \emph{IEEE Embedded Systems Letters},
  pp. 1--1, 2025.

\bibitem{he_sparse-tpu_2020}
X.~He, S.~Pal, A.~Amarnath, S.~Feng, D.-H. Park \emph{et~al.},
  ``\BIBforeignlanguage{en}{Sparse-{TPU}: adapting systolic arrays for sparse
  matrices},'' in \emph{\BIBforeignlanguage{en}{34th {ACM} {International}
  {Conference} on {Supercomputing}}}, Jun. 2020, pp. 1--12.

\bibitem{kiamari_gkan_2024}
M.~Kiamari, M.~Kiamari, and B.~Krishnamachari, ``{GKAN}: {Graph}
  {Kolmogorov}-{Arnold} {Networks},'' Jun. 2024, arXiv:2406.06470 [cs].

\bibitem{bodner_convolutional_2025}
A.~D. Bodner, A.~S. Tepsich, J.~N. Spolski, and S.~Pourteau, ``Convolutional
  {Kolmogorov}-{Arnold} {Networks},'' Mar. 2025, arXiv:2406.13155 [cs].

\bibitem{de1972calculating}
C.~De~Boor, ``On calculating with b-splines,'' \emph{Journal of Approximation
  theory}, vol.~6, no.~1, pp. 50--62, 1972.

\bibitem{integeronly}
B.~Jacob, S.~Kligys, B.~Chen, M.~Zhu, M.~Tang \emph{et~al.}, ``Quantization and
  training of neural networks for efficient integer-arithmetic-only
  inference,'' 2017, arXiv:1712.05877 [cs.LG].

\bibitem{bsplineprops}
T.~Lyche, C.~Manni, and H.~Speleers, \emph{Foundations of Spline Theory:
  B-Splines, Spline Approximation, and Hierarchical Refinement}.\hskip 1em plus
  0.5em minus 0.4em\relax Springer International Publishing, 2018, pp. 1--76.

\bibitem{dbb}
Z.-G. Liu, P.~N. Whatmough, Y.~Zhu, and M.~Mattina, ``S2ta: Exploiting
  structured sparsity for energy-efficient mobile cnn acceleration,'' 2022,
  arXiv:2107.07983 [cs.AR].

\bibitem{Kang_2020}
H.-J. Kang, ``Accelerator-aware pruning for convolutional neural networks,''
  \emph{IEEE Transactions on Circuits and Systems for Video Technology},
  vol.~30, no.~7, pp. 2093--2103, 2020.

\bibitem{sparsetc}
M.~Zhu, T.~Zhang, Z.~Gu, and Y.~Xie, ``Sparse tensor core: Algorithm and
  hardware co-design for vector-wise sparse neural networks on modern gpus,''
  in \emph{52nd Annual IEEE/ACM International Symposium on Microarchitecture
  (MICRO)}, ser. MICRO-52, 2019, p. 359–371.

\bibitem{cathelin_fully_2017}
A.~Cathelin, ``Fully {Depleted} {Silicon} on {Insulator} {Devices} {CMOS}:
  {The} 28-nm {Node} {Is} the {Perfect} {Technology} for {Analog}, {RF}, {mmW},
  and {Mixed}-{Signal} {System}-on-{Chip} {Integration},'' \emph{IEEE
  Solid-State Circuits Magazine}, vol.~9, no.~4, pp. 18--26, 2017.

\bibitem{fpmax}
J.~Pu, S.~Galal, X.~Yang, O.~Shacham, and M.~Horowitz, ``{FPMax: a 106GFLOPS/W
  at 217GFLOPS/mm2 Single-Precision FPU, and a 43.7GFLOPS/W at 74.6GFLOPS/mm2
  Double-Precision FPU, in 28nm UTBB FDSOI},'' 2016, arXiv:1606.07852 [cs.AR].

\bibitem{ucrtsdata}
H.~A. Dau, A.~Bagnall, K.~Kamgar, C.-C.~M. Yeh, Y.~Zhu \emph{et~al.}, ``The ucr
  time series archive,'' \emph{IEEE/CAA Journal of Automatica Sinica}, vol.~6,
  no.~6, pp. 1293--1305, 2019.

\bibitem{Ismail-Fawaz2023kan-c22-4-tsc}
A.~Ismail-Fawaz, M.~Devanne, S.~Berretti, J.~Weber, and G.~Forestier,
  ``Feature-based time series classification with kolmogorov–arnold
  networks,'' \url{https://github.com/MSD-IRIMAS/Simple-KAN-4-Time-Series},
  2024.

\bibitem{prefetching}
D.~Kulkarni, B.~Bhammar, H.~Thaker, P.~Dhobi, R.~P. Gohil \emph{et~al.}, ``A
  case for kolmogorov-arnold networks in prefetching: Towards low-latency,
  generalizable ml-based prefetchers,'' 2025, arXiv:2504.09074 [cs.AR].

\bibitem{efficientkan}
``An efficient implementation of kolmogorov-arnold network,''
  \url{https://github.com/Blealtan/efficient-kan}.

\bibitem{convkan}
V.~Starostin, ``Convolutional kan layer,''
  \url{https://github.com/StarostinV/convkan}, 2024.

\bibitem{mnist}
L.~Deng, ``The mnist database of handwritten digit images for machine learning
  research [best of the web],'' \emph{IEEE Signal Processing Magazine},
  vol.~29, no.~6, pp. 141--142, 2012.

\bibitem{cifar10}
A.~Krizhevsky, ``Learning multiple layers of features from tiny images,'' 2009.

\bibitem{drokin2024convkan}
I.~Drokin, ``Kolmogorov-arnold convolutions: Design principles and empirical
  studies,'' 2024, arXiv:2407.01092 [cs.AR].

\end{thebibliography}
\bibliographystyle{IEEEtran}

\end{document}